\documentclass[%
 reprint,
 amsmath,amssymb,
 aps,
]{revtex4-2}

\usepackage{amsmath}
\usepackage{graphicx}% Include figure files
\usepackage{dcolumn}% Align table columns on decimal point
\usepackage{bm}% bold math
\usepackage{hyperref}% add hypertext capabilities
\begin{document}

%\preprint{APS/123-QED}

\title{Practical limits for large-momentum-transfer clock atom interferometers}

\author{Mauro Chiarotti}
\author{Jonathan N. Tinsley}
\author{Satvika Bandarupally}
\author{Shamaila Manzoor}
\author{Michele Sacco}
\author{Leonardo Salvi}
\altaffiliation[Also at ]{INFN-Sezione di Firenze, Sesto Fiorentino, Italy}%
\author{Nicola Poli}%
\altaffiliation[Also at ]{INFN-Sezione di Firenze and CNR-INO, Sesto Fiorentino, Italy}%
\email[Author to whom correspondence should be addressed: ]{nicola.poli@unifi.it}%
\affiliation{Dipartimento di Fisica e Astronomia and LENS, Universit\`a degli Studi di Firenze, Via G. Sansone 1, 50019 -- Sesto Fiorentino, Italy}

\date{\today}% It is always \today, today,


%\title{Watt-level Hz-linewidth continuous UV laser source spectroscopy on the $^{1}$S$_{0}$~-~$^{3}$P$_{0}$ clock transition of cadmium}

%\title{Watt-level Hz-linewidth ultraviolet laser source for clock interferometry with cadmium atoms}

%\title{A Hz-linewidth high-power ultraviolet laser source for clock interferometry with cadmium atoms}

%%%%%%%%%%%%%%%    TWO COLUMN LETTER STYLE %%%%%%%%%%%%%%%%%%%%%%%%%%%%%%%%%%%%%%%

%\author[a]{Jonathan N. Tinsley}
%\author[a]{Satvika Bandarupally}
%\author[a]{Mauro Chiarotti}
%\author[a]{Shamaila Manzoor}
%\author[a,c]{Leonardo Salvi}
%\author[a,b,c]{Nicola Poli}
%%%%%%%%%%%%%%%%%%%%%%%%%%%%%%%%%%%%%%%%%%%%%%%%%%%%%%%%%%%%%%%%%%%%%%%%%%%%%%%%%

%%%%%%%%%%%%%%%%%%%%%%%%%%%%%%%%%%%%%%%%%%%%%%%%%%%%%%%%%%
%%% OTHER STYLE
%%%%%%%%%%%%%%%%%%%%%%%%%%%%%%%%%%%%%%%%%%%%%%%%%%%%%%%%%
%\author{Author One,\authormark{1,2,3}, Author Two \authormark{2} and Author Three \authormark{1,*}}

%\address{\authormark{1} Dipartimento di Fisica and LENS\\Universit\`a degli Studi di Firenze, Sesto Fiorentino, Italy\\
%\authormark{2}Istituto Nazionale di Ottica del Consiglio Nazionale delle Ricerche Sesto Fiorentino, Italy \\
%\authormark{3} Istituto Nazionale di Fisica Nucleare, Sezione di Firenze, Sesto Fiorentino, Italy}
%\email{\authormark{*}Corresponding author: nicola.poli@unifi.it} 

%\homepage{http://www.coldatoms.lens.unifi.it/poli} 


\begin{abstract}
Atom interferometry on optical clock transitions is being pursued for numerous long-baseline experiments both terrestrially and for future space missions. Crucial to meeting these experiments' required sensitivities is the implementation of large momentum transfer ($>$10$^3\hbar k$). Here we show that to sequentially apply such a large momentum via $\pi$ pulses places stringent requirements on the frequency noise of the interferometry laser, finding that the linewidth is required to be considerably lower than has previously been suggested. This is due to imperfect pulse fidelity in the presence of noise and is apparent even for an atom at rest interacting with resonant light, making this a fundamental constraint on operational fidelity for a given laser and pulse sequence. Within this framework, we further present and analyse two high-power, frequency-stabilised laser sources designed to perform interferometry on the $^1$S$_0$~-~$^3$P$_0$ clock transitions of cadmium and strontium, respectively operating at 332~nm and 698~nm.
\end{abstract}

\maketitle

%%%%%%%%%%%%%%%%%%%%%%%%%%  body  %%%%%%%%%%%%%%%%%%%%%%%%%%

\section{Introduction}
\label{sec:intro}
Atom interferometry utilising single-photon optical transitions represents an emerging technology with the ability to probe physics in a variety of previously untested regimes~\cite{Hu_2017,Hu_2019,Rudolph_2020}. This includes experiments looking for tests of the quantum twin paradox~\cite{Loriani2019}, gravitational red shift~\cite{Roura2020}, and tests of the weak equivalence for atoms in quantum superpositions~\cite{Rosi_2017}. In particular, however, multiple experiments based upon the $^1$S$_0$~-~$^3$P$_0$ clock transition of Sr at 698~nm have been proposed to search for a very wide set of fundamental physics goals, but especially for ultra-light dark matter searches and gravitational wave detection~\cite{Yu2011,Graham2013,SAGE_2019,AEDGE_2020,AION_2020,MAGIS_2021}. These experiments are predicated on the crucial fact that for interferometers based upon a single-photon transition, the laser phase is set at the point of emission and is therefore identical for all regions of the interferometer, regardless of their spatial separation, and cancels out in the read-out stage for a gradiometric configuration~\cite{Yu2011,Graham2013}. This allows for much larger interferometer baselines to be employed than would be possible with counter-propagating, multi-photon interferometry beams, as used in Raman or Bragg interferometers, where the finite speed of light introduces a noise term~\cite{LeGouet_2007}.

The use of large baselines and the enabled very large interferometry times are not, however, in of themselves sufficient to achieve the sensitivity required for the physics goals of these experiments. Therefore it is necessary to enhance the device sensitivity by employing the technique of large momentum transfer (LMT)~\cite{Graham2013}, with very large enhancements of 10$^3$-10$^4$ ultimately proposed for terrestrial experiments currently in the development stage~\cite{AION_2020,MAGIS_2021}. In practice, this typically means increasing the momentum separation between the two paths of the wavepacket by applying a series of $\pi$ pulses whose number is of the same order as the required enhancement~\cite{Graham2013,Rudolph_2020}. For these schemes it is therefore crucial that each $\pi$ pulse is as efficient as possible if interferometry contrast and fringe visibility is to be maintained, as a loss of fringe visibility leads to a decrease in device sensitivity and therefore acts in opposition to the metrological gain that would otherwise be achieved by LMT. From a practical standpoint, this requirement translates as needing an ultracold, velocity-selected atom source illuminated by an intense and homogeneous beam. Furthermore, although laser frequency noise cancels out in the phase readout in the clock-transition gradiometer configuration~\cite{Hu_2017,Hu_2019}, laser noise can nevertheless degrade device performance by reducing $\pi$ pulse efficiency and interferometer contrast and visibility~\cite{Szigeti_2012}, as has been experimentally observed for clock atom interferometers~\cite{Hu_2017,Hu_2019}.

In this article, we concentrate on the effect of laser noise, both intensity but especially frequency noise, on pulse fidelity and hence interferometer fringe visibility. We determine the fidelity for the Mach-Zehnder interferometry schemes proposed for large-baseline clock-transition interferometers~\cite{Graham2013}, showing that the induced imperfections of fidelity lead to practical difficulties in scaling LMT to the desired levels of large-baseline clock atom interferometers, especially at the proposed laser linewidth of $\sim$10~Hz~\cite{MAGIS_2021}. %The resulting loss of contrast is shown to lead to the enhancement of interferometer sensitivity ceasing to be linear with LMT and an optimum LMT factor exists for a given configuration. 
It is further shown that, in general, the fidelity of the pulse sequence is improved for a given noise level with increasing Rabi frequency, highlighting the need for a high-intensity interferometry beam, which is an important technological difference between using optical clock transitions for atom interferometry and frequency metrology. As we consider the case of a single two-level atom interacting with a resonant laser, these results hold regardless of the atom cloud temperature or spatial spread and represent the maximum achievable fidelity for a given laser system and pulse sequence.

We further apply this fidelity formalism to two newly developed high-power Hz-level clock lasers and show the practical LMT limits in principle achievable with these lasers when performing a standard clock interferometry sequence. These two systems have been specifically developed for performing atom intererometry on the $^1$S$_0$~-~$^3$P$_0$ clock transitions of Cd and Sr at 332~nm and 698~nm, respectively~\cite{Tinsley_2022}. For the case of the UV transition of Cd, we build upon on recently demonstrated technology for a laser source centred at 326~nm~\cite{Manzoor_2022}, extending that work by demonstrating the capability of obtaining high coherence via successive frequency-stabilisation stages on optical cavities with increasing finesse. 

The structure of this article is the following: in Section~\ref{sec:lmtContrast} we discuss the basics of LMT clock interferometry and visibility loss; in Section~\ref{sec:pulseFidelity} we present the basic theory of pulse fidelity, with further details in Appendix~\ref{app:filterFunction}, and show the results of the pulse fidelity calculations applied to an LMT sequence; Section~\ref{sec:lasersource} discusses the developed laser systems and analyses them within this framework; and conclusions are reported in Section~\ref{sec:summary}.

%Here we follow such a formalism for pulse fidelity~\cite{Green_2013} to determine the fidelity for the Mach-Zehnder interferometry schemes proposed for large-baseline clock-transition interferometers~\cite{Graham2013}. Following this formalism, it is shown that there is a rapid decrease in total fidelity for an increasing number of $\pi$ pulses and that the LMT goals of these experiments cannot be achieved with a 10~Hz laser operating at the proposed Rabi frequencies and using a standard pulse sequence. It is further shown that, in general, the pulse sequence fidelity is improved for a given noise level with increasing Rabi frequency, highlighting the need for a high-intensity interferometry beam, which is an important technological difference between using optical clock transitions for atom interferometry and frequency metrology. As we consider the case of a single two-level atom interacting with a resonant laser, these results hold regardless of the atom cloud temperature or spatial spread.

\section{LMT Clock Atom Interferometer Sensitivity}\label{sec:lmtContrast}
\begin{figure}[t]
\centering\includegraphics[width=8.6cm]{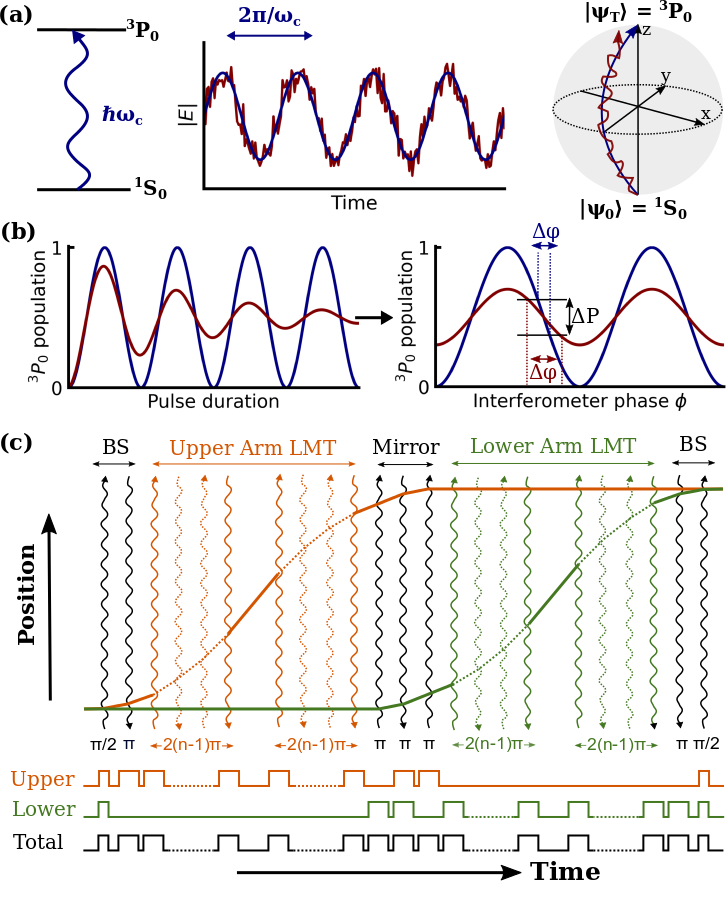}
\caption{(a) Clock atom interferometers operate on the metastable $^1$S$_0$~-~$^3$P$_0$ transition, which is ideally driven by a noiseless laser (blue line). A noisy laser system (red line) will instead introduce errors when applying operations, such as for a $\pi$ pulse as shown here in the Bloch sphere representation of the two-level system. (b) These imperfections lead to damping in the temporal domain and also to a loss of interferometry visibility and hence interferometric sensitivity. (c) Proposed LMT scheme for a large-area, clock atom interferometer~\cite{Graham2013}. Black lines represent the basic interferometry scheme, whilst orange and green lines represent the LMT pulses applied to the upper and lower arms, respectively. The upper and lower arms of the interferometer receive different pulses, but with a pattern which is symmetric in time.}
\label{fig:lmtSetup}
\end{figure}

The leading-order phase shift of a clock atom interferometer is $\phi\sim nkgT^2$, where $n$ is the LMT order, $k$ is the wavenumber of the transition, $g$ is the local gravitational acceleration, and $T$ the interferometry time~\cite{Hu_2017,Graham2013}. Single-shot interferometry sensitivity can be characterised by the ratio $\Delta\phi/\phi$ where $\Delta\phi$ is the phase difference between the two interferometer arms that can be experimentally resolved. It is therefore expected that, in principle, device sensitivity enhancement should scale linearly with LMT order $n$ and it has consequently been suggested to improve clock atom interferometers by performing a succession of $\pi$ pulses propagating in alternating directions~\cite{Graham2013}. In brief, an LMT enhancement of $n$ is achieved by applying two additional stages of $2(n-1)$ $\pi$ pulses independently resonant to the upper and lower interferometry arms and temporally mirrored about the standard interferometry mirror pulse sequence (Fig.~\ref{fig:lmtSetup}). 

%However, $\Delta\phi$ is not itself necessarily independent of the parameters $n$ and $T$ as it depends upon interferometer contrast.
However, in experimental conditions $\Delta\phi$ itself is typically found to be dependent upon the parameters $n$ and $T$. This is because $\Delta\phi$ is extracted from the relative population difference and is therefore proportional to the slope and hence the visibility of the interferometry fringe (see Fig.~\ref{fig:lmtSetup}). In a typical atom interferometer, sources of visibility loss relating to the interferometry pulses can be roughly divided into two categories: those which lead to the different constituent atoms of the ensemble becoming out of phase with each other; and those which degrade the operational fidelity on the single-atom scale~\cite{Green_2013}. The former is well-studied and often represents the leading practical constraint on atom interferometers, with finite atom cloud temperature and size, as well as interferometry beam imperfections (e.g. wavefront and intensity inhomogeneities), causing different atoms within the ensemble to interact with effectively different laser beams during the sequence~\cite{Szigeti_2012,Louchet_Chauvet_2011,Hilico_2015}. The resulting atomic dephasing causes a loss of contrast and visibility, as the read-out of the interferometer is effectively the average over all the individual atoms. Not all such contrast losses, however, can necessarily be considered as due to a fundamental loss of coherence, as demonstrated by related spin-echo techniques~\cite{Hahn_1950,Solaro_2016}. Nevertheless, such effects are known to be both crucial and a key limiting factor for many current atom interferometers~\cite{Szigeti_2012}.

In addition to beam and cloud temperature imperfections reducing visibility, however, imperfect application of the desired quantum operation (e.g. $\pi/2$ and $\pi$ pulses) will lead to contrast loss and other systematic errors. One source of such imperfections is laser noise and crucially LMT exacerbates the problem as it increases the number of necessarily imperfect pulses that are applied. In investigating this problem, previous analyses have tended to account for laser frequency noise by considering the excited population fraction from a $\pi$ pulse to be $P_e\approx1-\delta^2/\Omega_R^2$~\cite{Graham2013,MAGIS_2021}, with $\delta$ the effective detuning and $\Omega_R$ the Rabi frequency. In this regime, it is the small changes in $\delta$ caused by the non-zero laser linewidth that are considered. Such calculations have concluded that, for $\Omega_R\sim$10$^3$~rad/s, the laser linewidth needs to be kept to the order of 10~Hz for contrast loss on the percent level for an interferometer with LMT order $n\sim$10$^3$~\cite{MAGIS_2021}. As 10~Hz is readily achievable with current technologies, other effects such as finite atom temperature and intensity variations across the atom cloud are considered as the main source of dephasing and thus contrast and visibility loss.

However, it is also well known that the laser noise in terms of both intensity~\cite{Thom_2018} and, especially, frequency~\cite{Ball_2016,Levine_2018} is an important parameter in determining pulse fidelity in quantum systems, as has been extensively studied in the field of quantum computing which targets 10$^4$ operations on a single qubit system~\cite{Knill_2005,Gaebler_2016}. Furthermore, it has recently been shown that to correctly determine $\pi$-pulse fidelity it is not sufficient to consider only the linewidth of the interrogating laser, but rather the whole power spectrum of fluctuations should be considered~\cite{Day_2021}. Imperfect pulse fidelity arising from laser noise reduces the visibility as above, but in an arguably more fundamental way as it is a damping of a single atom rather than a washing-out across the ensemble.

We note that although this laser noise also affects the fidelity of e.g. Raman transitions~\cite{Day_2021}, this represents a crucial practical difference between clock atom interferometers and more standard, multi-photon transition configurations. In the latter case the relevant noise term is the \emph{relative} phase noise between the counter-propagating interferometry beams, whilst for clock interferometers it is the \emph{absolute} laser phase noise itself which is the relevant parameter~\cite{Hu_2019}. The phase noise of Raman and Bragg beams can be easily reduced to orders of magnitude less than the phase noise that is currently achievable with even the narrowest clock lasers~\cite{Matei_2017,Zhang_2017}.

\section{Pulse Fidelity \& Computational Results}\label{sec:pulseFidelity}

\begin{figure*}[t]
\centering\includegraphics[width=\textwidth]{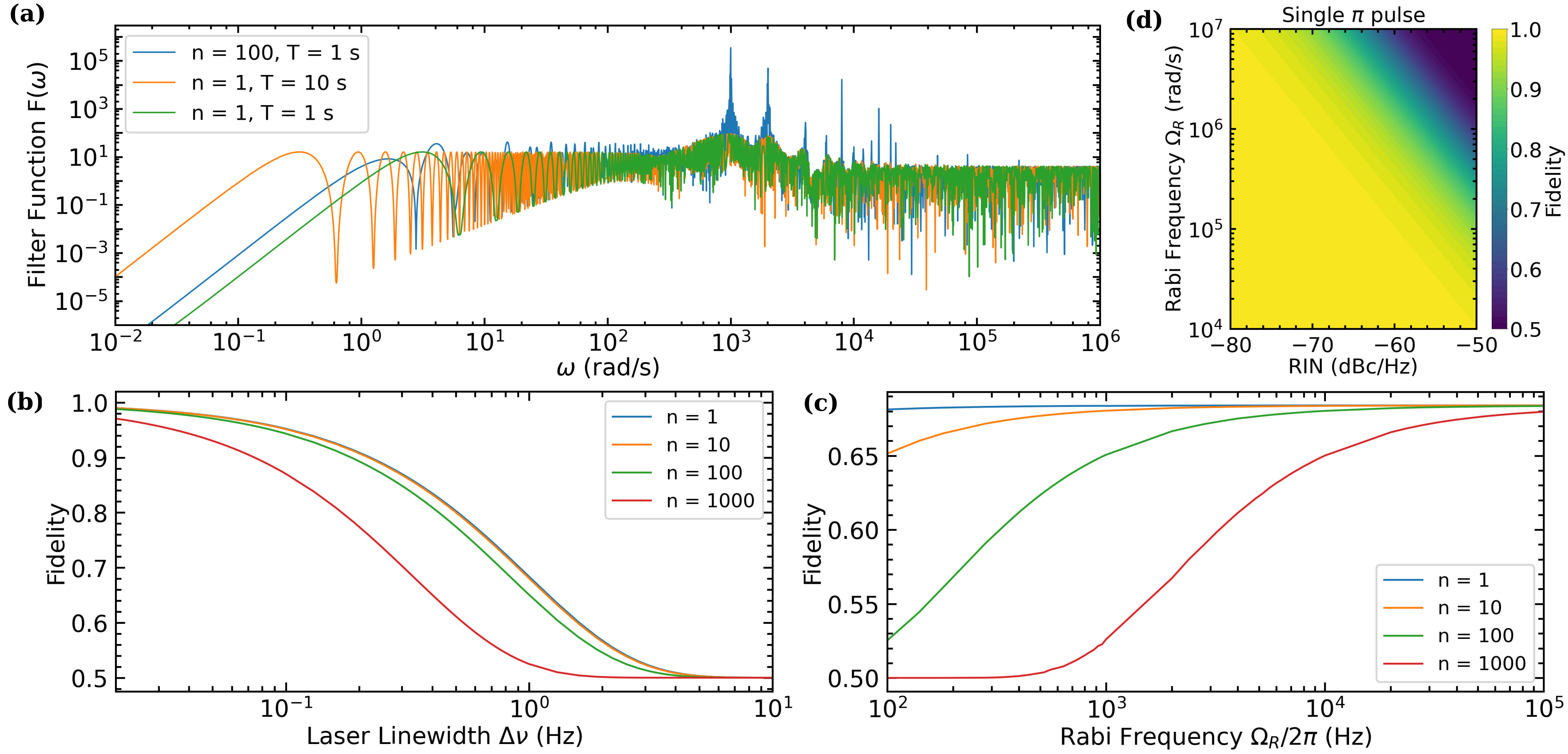}
\caption{(a) Filter function calculated for different LMT values (see Fig.~\ref{fig:lmtSetup}(a)). In these calculations, $T_1$=1~s or 10~s and $T_1\simeq T_2$ due to the high Rabi frequency values ($\Omega_R$=10$^3$~rad/s) and short pulse separation ($\tau$=1~$\mu$s). Fidelity of an LMT sequence as a function of (b) laser linewidth for a fixed Rabi frquency ($\Omega_R = 2\pi\times10^3$~Hz) and (c) as a function of Rabi frequency at a fixed laser linewidth ($\Delta\nu$=1~Hz). In both cases, $\tau$=1~$\mu$s and $T_1$=1~s. d) Fidelity of a single $\pi$ pulse as a function of RIN level and Rabi frequency.}
\label{fig:fidelity}
\end{figure*}

In determining the effect of the laser noise, we consider the fidelity of the whole interferometry sequence, where the fidelity of a quantum operation is defined as the overlap of the target operation with the actually applied operation. This is quantified by the  square modulus of the Hilbert-Schmidt inner product of these operations~\cite{Green_2013}. %target state $|\psi_T\rangle$ and the state produced by the operation $\hat{U}$ on the initial state $|\psi_0\rangle$: $\mathcal{F}=\left|\langle\psi_T|\hat{U}|\psi_0\rangle\right|^2$. For a $\pi$ pulse operating on the ground state, the fidelity is therefore equivalent to the population transfer to the excited state.
Laser noise causes the state evolution on the Bloch sphere, defined here by the two-level system formed by the $^1$S$_0$~-~$^3$P$_0$ clock transition, to deviate from the ideal case and thus for the fidelity to deviate from unity (Fig.~\ref{fig:lmtSetup}). Specifically, in the standard case of applying a rotation about the x or y axes, frequency and intensity noise respectively produce undesired rotations about the $z$ axis and the $x$ and $y$ axes~\cite{Day_2021}. In the presence of weak noise, the fidelity can be written as~\cite{Green_2013,Ball_2016,Day_2021},
\begin{equation} \label{eq:fidelity}
    \mathcal{F}\simeq\frac{1}{2}\left[1+e^{-\chi}\right],
\end{equation}
where $\chi$ is the fidelity decay constant which is related, to first-order, to the noise power spectral density (PSD) by the filter function ($F\left(\omega\right)$) of the operation being performed:
\begin{equation} \label{eq:chi}
    \chi=\sum_j\frac{1}{\pi}\int^{\infty}_0\frac{d\omega}{\omega^2}S_j\left(\omega\right)F_j\left(\omega\right);
\end{equation}
where $S_j\left(\omega\right)$ is linearly related to the PSD in a manner dependent upon the type of noise source and the summation is over the different sources of noise. Specifically, for frequency and intensity noise $S_j\left(\omega\right)=\frac{1}{4}S_{\omega_{L}}\left(\omega\right)$ and $S_j\left(\omega\right)=\frac{1}{16}\Omega_R^2S_{I}\left(\omega\right)$, respectively, where $S_{\omega_{L}}\left(\omega\right)$ is the laser frequency noise, $S_{I}\left(\omega\right)$ the relative intensity noise (RIN), and $\Omega_R$ is the Rabi frequency~\cite{Day_2021}. The problem of determining the fidelity of a quantum operation is therefore reduced to the knowledge of the relevant PSDs and calculating the appropriate filter functions.

\begin{figure}[t]
\centering\includegraphics[width=8.6cm]{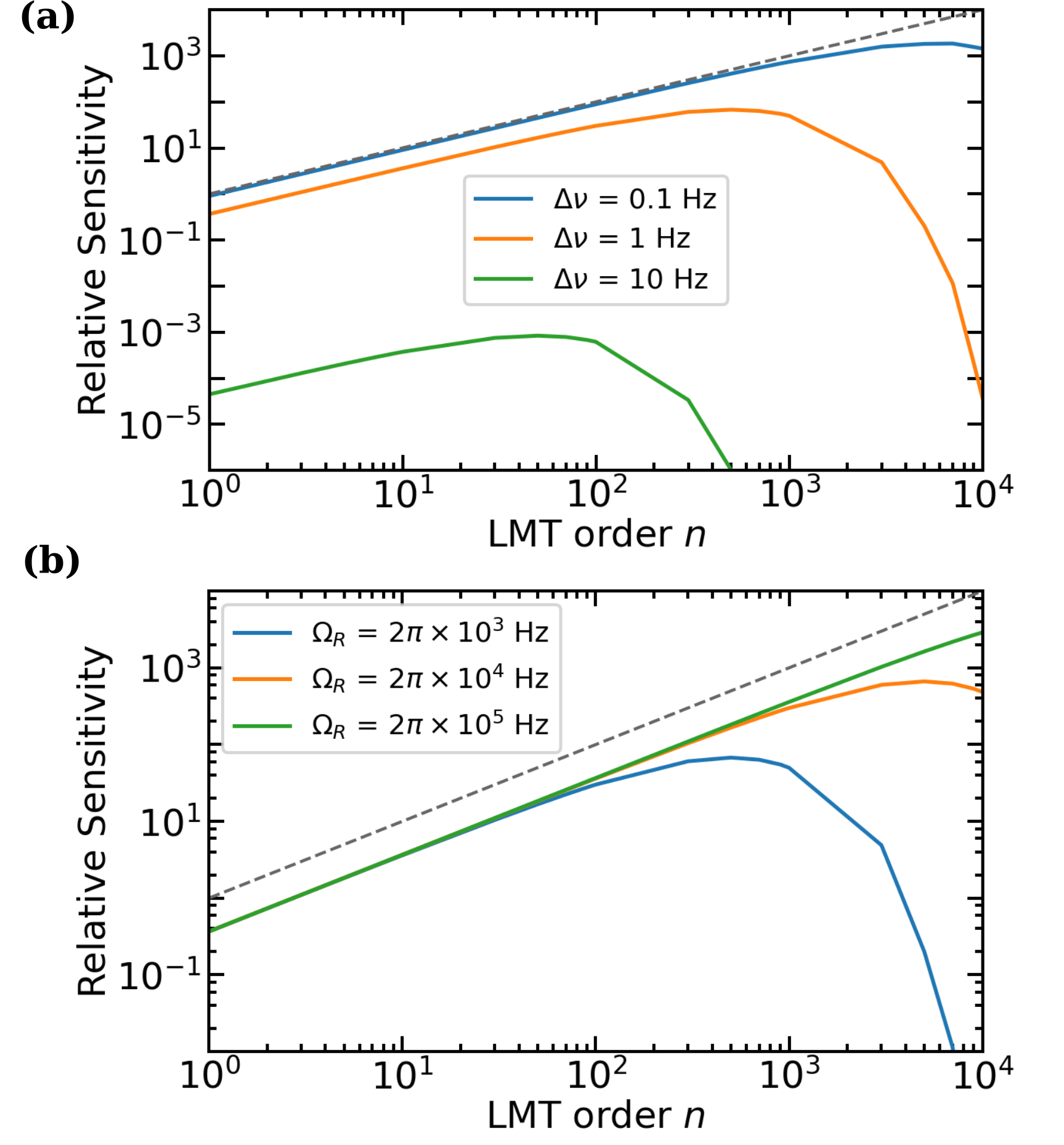}
\caption{Relative interferometer sensitivity, defined as $nW$, as a function of LMT for different values of (a) laser linewidth $\Delta\nu$ (with $\Omega_R=2\pi\times10^3$~Hz) and (b) Rabi frequency $\Omega_R$ (with $\Delta\nu$=1~Hz). Dashed grey lines show linear scaling with LMT and $W=1$ for comparison. In all cases, $\tau$=1~$\mu$s and $T_1$=1~s.}
\label{fig:relSenstivityFig}
\end{figure}

We determine the filter function for laser frequency noise for a sequence of pulses mimicking the proposed LMT sequence in a clock-transition interferometer for detecting gravitational waves~\cite{Graham2013} (Fig.~\ref{fig:lmtSetup}(c)). These calculations follow a known method~\cite{Green_2013}, with the key results for this application summarised in Appendix~\ref{app:filterFunction}. Figure~\ref{fig:fidelity} shows the filter function calculated for a variety of LMT values and shows a structure with a low-frequency cut-on, two principle resonances around $\Omega_R$, and then a flatter distribution at higher frequencies with some harmonics on top. As can be seen, the filter function structure is largely unmodified with increasing LMT, except that the function resonances about the Rabi frequency are amplified approximately quadratically with LMT order $n$. Increasing the interferometry time, increases the sensitivity to low-frequency noise and determines the position of the function's nodes at low frequency. %In the case of a single $\pi$ pulse, this resonance has a value of 10 and is centred at $\left(2e^{-1}+1\right)\Omega_R\sim1.74\Omega_R$. For the LMT clock atom interferometry sequence these resonances appear at $\Omega_R$ and its even multiples. Increasing the interferometry time, increases the sensitivity to low-frequency noise.

The above framework can be used to compute the effect of the laser frequency noise and RIN on operational fidelity. We calculate the effects independently, but in the real case they must be combined through the summation in Eq.~\ref{eq:chi}. When analysing these effects we in all cases assume a spectrally flat distribution, i.e. the PSD is a constant value, and integrate Eq.~\ref{eq:chi} in the range 10$^{-2}$--10$^7$~rad/s. Although ideally Eq.~\ref{eq:chi} should be integrated over all angular frequencies, we find that, in our case, when increasing or decreasing both the upper and lower integration limits by two decades, the resultant fidelity is only altered by $\sim$1\%.

In the case of the RIN, it is observed that the pulse fidelity from a single $\pi$ pulse is relatively unaffected by the laser noise (Fig.~\ref{fig:fidelity}(d)), with excellent fidelities above 0.999 achievable even with high RIN values of -70~dBc/Hz for a wide range of Rabi frequencies. However, it is important to note that the $\Omega_R^2$ dependence in Eq.~\ref{eq:chi} may make the RIN contribution experimentally important should alternative schemes to increase the Rabi frequency be used~\cite{Rudolph_2020,Nourshargh_2021}. Nevertheless, we therefore ignore the contribution of the RIN in the following analysis.

Conversely, however, the fidelity of an LMT sequence is highly degraded by the presence of laser noise, unless the laser noise is reduced to experimentally difficult linewidths (Fig.~\ref{fig:fidelity}(b)). As mentioned, these calculations assume white noise which we relate to a Lorentzian linewidth via $\Delta\nu=\pi S_\nu=S_{\omega_{L}}/2$. We have focused on Lorentzian linewidths for generality, whereas in reality laser linewidths typically take the form of a central Gaussian region and a Lorentzian component at higher frequencies~\cite{DiDomenico_2010}, with an additional servo bump from the employed feedback circuit. The effect of these spectral features has been studied in a quantum computing context~\cite{Day_2021}, and allows for the possibility to potentially reduce the degrading effect of the laser noise on the interferometry fidelity by suppressing the noise around peaks of the filter function or by judicious selection of the Rabi frequency. Therefore, for any real application, the measured noise spectra of the laser sources should therefore be used for analysis, as we show later in Fig.~\ref{fig:clockLaserFigure}.

As an example of the difficulties posed, however, the fidelity already approaches its minimum value of 0.5 for $n=1000$ even with a linewidth of 1~Hz (Fig.~\ref{fig:fidelity}(b)). For a given LMT and $\Omega_R$, the fidelity is initially relatively unaffected by increasing $\Delta\nu$, but decays exponentially following a cut-off linewidth. In this regime, the fidelity decays from approximately 0.9 to the minimum 0.5 with an approximate factor 10 increase in $\Delta\nu$, meaning it is important to keep $\Delta\nu$ at or lower than this cut-off value to achieve the highest fidelities.

Unlike in the case of the RIN, however, the situation is improved with increasing Rabi frequency (Fig.~\ref{fig:fidelity}(c)), as this increases the frequency of the resonances (Fig.~\ref{fig:lmtSetup}(b)). Combined with the $1/\omega^2$ dependence of Eq.~\ref{eq:chi}, this leads to a relative suppression of the laser noise. This is in agreement with the simple picture of considering the ratio of the effective detuning and the Rabi frequency.  Increasing $\Omega_R$ has additional benefits for real systems with finite temperature~\cite{Nourshargh_2021}, although the RIN will become increasingly important, as discussed above.

To quantify the reduction in interferometer performance, we consider the parameter $W=\text{exp}\left[{-\chi}\right]$ which varies from 0--1 and is closely related to the fidelity (Eqs.~\ref{eq:fidelity} and~\ref{eq:chi}). For a standard Mach-Zehnder sequence, this $W$ parameter scales linearly with the interferometer visibility. Therefore assuming that the interferometer sensitivity is linearly dependent upon both this value and LMT order (see Sec.~\ref{sec:lmtContrast}), this allows for the sensitivity relative to an interferometer with $W=1$ and $n=1$ to be determined. In the ideal case of perfect pulse fidelity, this relative sensitivity enhancement would scale linearly with $n$, but this is seen to not be the case in the presence of laser frequency noise. As shown in Fig.~\ref{fig:relSenstivityFig}, there is in fact an optimal value of $n$ for a given $\Delta\nu$ and $\Omega_R$ in terms of maximising single-shot sensitivity. For example, for a laser with $\Delta\nu$=1~Hz and $\Omega_R = 2\pi\times10^3$~Hz this occurs for $n\sim500$ and at $n$=10$^4$ the sensitivity is lower than at $n$=1 (Fig.~\ref{fig:relSenstivityFig}(b)). For a 10-Hz-linewidth laser at the same Rabi frequency, however, this optimum is at $n<100$ and, moreover, the relative sensitivity is significantly less than 1 in all cases (Fig.~\ref{fig:relSenstivityFig}(a)), highlighting the necessity of low-noise lasers if the potential sensitivity gains from LMT are to be realised.

%However, as at high values of LMT the Mach-Zehnder sequence becomes dominated by the LMT $\pi$ pulse sequence, we also consider the simpler case of a sequence of $\pi$ pulses only, with the re lationship between LMT order and the total number of $\pi$ pulses being.... This makes a negligible difference to the filter function, meaning that the large separation between the two stages of the $\pi$ pulses in the Mach-Zehnder configuration does not play a key role (Fig.~\ref{fig:lmtSetup}). When considering only the case of $\pi$ pulses we consider also the factor $W=\left[1+e^{-\chi}\right]$, which can be related to the interferometer contrast~\cite{Szigeti_2012}.

\section{Clock Interferometry Laser Sources}
\label{sec:lasersource}
\begin{figure}[tb]
\centering\includegraphics[width=8.6cm]{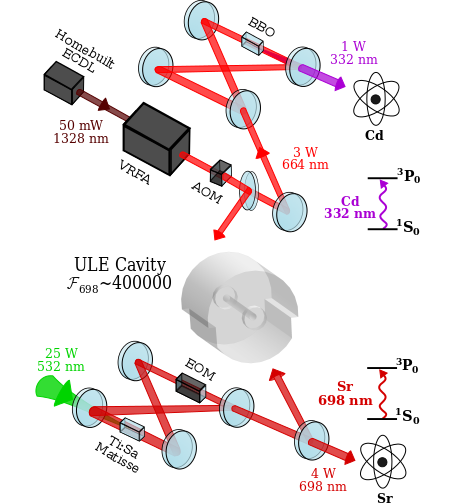}
\caption{Setup of the clock atom interferometry laser sources for Sr and Cd at 332~nm and 698~nm, respectively. The laser for Sr is derived from a Ti:sapphire laser with an intra-cavity EOM. For Cd, the laser is a frequency-quadrupled ECDL which is amplified by a Raman fibre amplifier. The lasers can both be locked to the same super-high-finesse ULE clock cavity. See text for details of the locking procedures.}
\label{fig:laserSetup}
\end{figure}

\begin{figure*}[htbp]
\centering\includegraphics[width=\textwidth]{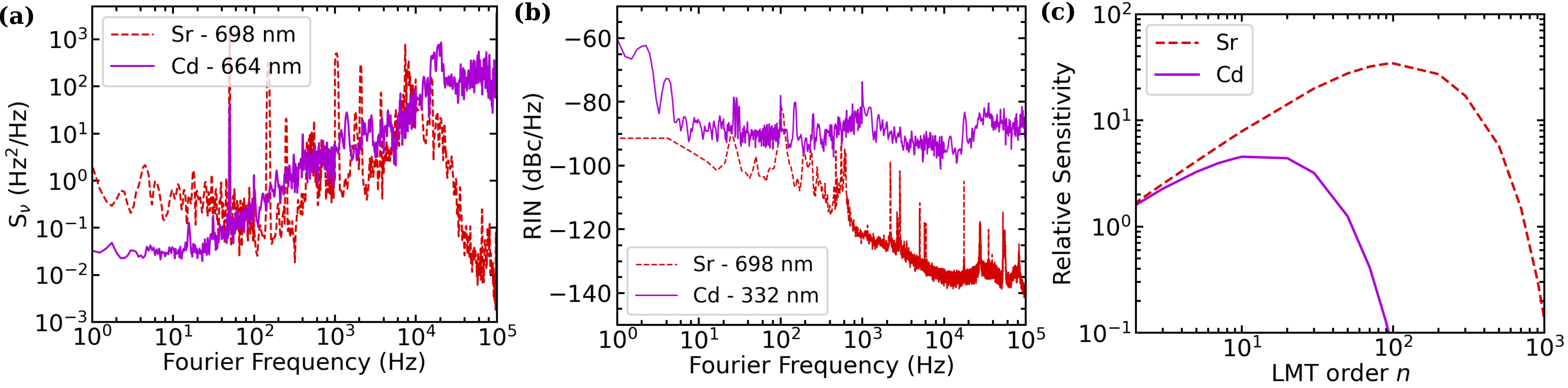}
\caption{(a) Measured in-loop $S_\nu\left(f\right)$ for the two clock interferometry lasers at 698~nm and 664~nm for Sr and Cd, respectively. (b) Free-running RIN of the two clock interferometry lasers. (c) Relative interferometer sensitivity as a function of LMT for for the developed laser sources with $\tau$=1~$\mu$s and $T$=100~ms and $\Omega_\text{Sr}=2\pi\times$1.7~kHz and $\Omega_\text{Cd}=2\pi\times$740~Hz. In all plots dashed red lines are for Sr and solid purple lines are for Cd.}
\label{fig:clockLaserFigure}
\end{figure*}

As can be deduced from the above results, the principle requirements for a laser system for performing LMT clock-transition interferometry are high optical power and high coherence (Fig.~\ref{fig:laserSetup}). For the Sr system the $^1$S$_0$~-~$^3$P$_0$ is at 698~nm meaning that commercial systems can provide the basis of this, in addition to providing good spatial mode quality and low RIN. We use a commercial Ti:sapphire system (Matisse TS, Sirah), pumped with up to 25~W at 532~nm, which is locked to a super-high-finesse ultra low expansion (ULE) optical Fabry-P\'{e}rot cavity with a finesse $\mathcal{F}_{698}=4.1\times10^{5}$~\cite{Tarallo_2011}. A high feedback bandwidth ($\sim$300~kHz) is achieved by utilising an intra-cavity electro-optical modulator (EOM), at the cost of a slight reduction in output power ($\sim$20$\%$). Nevertheless, a total optical power of 4~W is achieved at 698~nm in this configuration. Stabilisation is performed by pre-stabilising the laser to a medium-finesse Fabry-P\'{e}rot cavity, using the EOM and the piezoelectric transducers (PZTs) on two of the laser mirrors. This cavity is in turn locked to the super-high-finesse cavity by controlling its length with a PZT on one of the cavity mirrors. For both cavities $\sim$1~mW of light is sent via polarisation-maintaining fibre and modulated at $\sim$10~MHz by external EOMs for Pound-Drever-Hall locking (PDH)~\cite{Drever_1983}. The PSD of the in-loop measurement of the error signal is shown in Fig.~\ref{fig:clockLaserFigure} and yields a linewidth of 10~Hz.

The $^1$S$_0$~-~$^3$P$_0$ of Cd, however, is at 332~nm and a UV light source capable of 1~W of optical power and coherence at the Hz level is not currently commercially available. It was therefore necessary to develop a system internally, optimising its performance ad hoc for this application. We follow the same basic scheme as was recently reported for a laser for the $^1$S$_0$~-~$^3$P$_1$ Cd transition at 326~nm~\cite{Manzoor_2022}, where intense UV radiation is produced by two successive stages of second-harmonic generation (SHG), i.e., it is quadrupled from the infrared emission of a high-power semiconductor diode and fibre amplifier. In brief, a Littrow-configuration extended-cavity diode master laser (ECDL) based on a gain chip is used to inject a commercial Raman fibre amplifier, which amplifies the 35-mW input to the 10-W level. SHG of this light is performed in a periodically-poled lithium niobate (PPLN) crystal, producing around 3~W at 664~nm, before being doubled again in a resonant bow-tie cavity using a Brewster-cut beta-barium borate ($\beta$-BaB$_{2}$O$_{4}$, BBO) crystal as the non-linear, doubling medium~\cite{Hannig_2018}. This cavity is locked via the H\"{a}nsch-Couillaud technique~\cite{Hansch_1980} for total produced UV powers up to $\sim$1~W.

The key difference with the previous scheme is the requirement of greater coherence than was necessary for the $^1$S$_0$~-~$^3$P$_1$ Cd transition ($\Gamma$=$2\pi\times$67~kHz). This is achieved first by optimising the pre-stabilisation lock of the ECDL onto a medium-finesse optical Fabry-P\'{e}rot cavity ($\mathcal{F}_{pre}=5\times10^{3}$). The stabilisation loop is closed by sending in feedback the error signal thus generated, after engineering through the use of optimised PIDs and filters, to two transduction channels: the PZT which in the ECDL defines the grating orientation (low frequency – 1~kHz) and the modulation input superimposed on the current supply (high frequency – 700~kHz). The in-loop noise analysis conducted on the described stage, which is capable of reducing the source noise to equivalence to an estimated linewidth of $\sim$Hz in the 100~kHz band~\cite{Halford_1971}.

To really reduce the linewidth further to the Hz level the laser must be locked to the ULE cavity, for which a second feedback stage is applied in the visible region following SHG. The observation of a 20~Hz beat between the pre-stabilised and the amplified radiation allows us to conclude that the coherence length of the source is so far sufficiently large to neglect the frequency noise introduced by the Raman fibre amplification stage. We also observe the presence of low-frequency noise (10~dB at 100~Hz, 30~dB at 5~kHz with respect to the central emission peak), to which this second stabilisation stage is directed. We use the same cavity and setup as for the Sr laser, with the only difference being a reduction in cavity finesse to $\mathcal{F}_{664}=1.3\times10^{5}$, as measured by the ring-down method. The PDH error signal is split and the low-frequency component is sent to the PZT of the pre-stabilisation cavity, to compensate drifts. An acousto-optical modulator (AOM), in single-pass on the visible SHG, is used for higher frequencies up to around 50~kHz. The AOM is also used to additionally stabilise the intensity and reduce the amplitude noise added by the amplification stage~\cite{Manzoor_2022}. With this method, the in-loop measured linewidth is 2~Hz at 664~nm, which will correspond to around 8~Hz after the doubling stage to the UV (Fig.~\ref{fig:clockLaserFigure}(a)). This value represents the lower limit for the linewidth, although the main noise contribution of the doubling stage should be frequency-to-amplitude conversion. Direct experimental validation of the linewidth or stablisation at 332~nm itself would require an evacuated high-finesse cavity in the UV regime, for which the production of the necessary substrates and coatings remains an active area of research~\cite{Burkley_2021}.

Finally, the radiation at 664~nm is coupled (through a T=1.5$\%$ input mirror) with a commercial bow-tie optical cavity, specifically designed for the application~\cite{Hannig_2018}. A production of UV power up to 1.2~W is achieved, which matches well with the expected values for the measured finesse $\mathcal{F}_{BBO} = 323 \pm 3$ and the nonlinear conversion coefficient $E_{nl} = 1.21 \times 10^{-4}$ calculated from the nominal system parameters~\cite{Pizzocaro_2014}. The RIN of the UV (Fig.~\ref{fig:clockLaserFigure}b) is kept low ($\sim$-90~dBc/Hz for 10-10$^5$~Hz) by the AOM stabilisation stage and by passively shielding the BBO cavity to reduce acoustic resonances and no further active stabilisation is required.

The measured intensity and frequency noise of these laser sources can be used to analyse their expected performance in an LMT clock atom interferometer. We input these measured values into Eq.~\ref{eq:chi} and integrate between the measurement range of 1--10$^5$~Hz. Inputting the measured RIN values for our two laser sources (Fig.~\ref{fig:clockLaserFigure}(b)) gives single-pulse fidelities of $\mathcal{F_\pi}>$0.9999 for $\Omega_R<10^6$ in both cases, suggesting the RIN will pose negligible problems as expected. To instead estimate the effect of the frequency noise, however, we first fix the expected Rabi frequency. For a Gaussian beam radius of 5~mm, which is well below the maximum permissible size to keep the intensity ripple at the 1\% level for diffraction from the standard DN100CF viewports which are to be employed~\cite{Siegman_1986}, we estimate achievable $\Omega_R$ values of $2\pi\times$1.7~kHz and $2\pi\times$740~Hz for the Sr and Cd sources, respectively. These Rabi frequencies can be used to estimate optimal LMT parameters of $n_{Sr}\sim$100 and $n_{Cd}\sim$10 for Sr and Cd, respectively (Fig.~\ref{fig:clockLaserFigure}(c)), based on the measured frequency noise (Fig.~\ref{fig:clockLaserFigure}(a)) and $T$=100~ms, which is a reasonable interferometry time for our proposed experiments~\cite{Tinsley_2022}. In the case of Sr, such an LMT-based enhancement in performance represents an in principle improvement upon what has previously been demonstrated with clock atom interferometry schemes~\cite{Rudolph_2020}.

\section{Conclusion \& Outlook}
\label{sec:summary}
We have provided a general analysis for the noise requirements for a system to perform LMT clock atom interferometry with a standard pulse sequence by investigating the operational fidelity in the single-atom case. These analyses have shown the challenging nature of proposed long-baseline experiments for e.g. gravitational wave detection and the difficulties in performing clock atom interferometer with LMT parameters up to $10^4\hbar k$. In order to fully understand the effect on interferometer performance, especially in a gradiometer configuration where some common-mode noise cancellation may persist, it would be necessary to extend the above calculations to a full quantum simulation incorporating parameters such as atom cloud temperature and spatial spread, loss mechanisms, and the external degrees-of-freedom of the system. Nevertheless, the outlined formalism will be useful for determining the interferometry laser requirements for the future experiments both on Earth~\cite{AION_2020,MAGIS_2021} and in space~\cite{SAGE_2019,AEDGE_2020} which seek to use clock atom interferometry to perform fundamental physics experiments. The maintenance of contrast and visibility is also crucial for achieving metrological gain in experiments which seek to use squeezing to go below the quantum projection noise limit~\cite{Pedrozo_2020,Greve_2021}.

The analysis presented above has further focused solely on square pulses and we note that it is possible to enhance $\pi$-pulse population transfer efficiency by techniques such as adiabatic rapid passage~\cite{Kovachy_2012,Kotru_2015} or composite pulses~\cite{Butts_2013,Dunning_2014}, which can further be optimised by pulse shaping~\cite{Saywell_2020,Saywell_2020_thesis,Chen_2022}. Alternative schemes, such as using the $^1$S$_0$~-~$^3$P$_1$ transition for the LMT portion of the interferometer~\cite{Rudolph_2020} or using a cavity to enhance the optical power and increase the Rabi frequency~\cite{Nourshargh_2021}, have also been previously been discussed and may help to circumnavigate some of these problems. Trapped-atom configurations which use Bloch oscillations to enhance the interferometry sensitivity may also offer a route towards a solution~\cite{Zhang_2016,Xu_2019,McAlpine_2020}. In all cases, however, our analysis suggests that the noise of the interferometry laser will play an important role and should not be discounted, but rather integrated into these approaches.

Within this operational fidelity framework, we have presented and analysed two laser systems suitable for performing clock atom interferometry -- one for the 698~nm transition of Sr and one for the 332~nm transition of Cd -- possessing the required characteristics of high power, low RIN, and $\sim$Hz-level linewidth. In the case of the 332-nm laser we have established the capability to produce high-coherence and high-power light in the UV by frequency quadrupling and amplifying an ECDL at 1328~nm. These lasers will be implemented in a dual-species atom interferometer, with potential applications to tests of the weak equivalence principle and quantum time dilation~\cite{Tinsley_2022}.

\begin{acknowledgments}
We thank Grzegorz D. Pekala for assistance in establishing the Cd clock laser source. We thank Guglielmo M. Tino and Hidetoshi Katori for useful discussions and Michael Holynski for a critical reading of the manuscript. This work has been supported by the European Research Council, Grant No.772126 (TICTOCGRAV).
\end{acknowledgments}

\appendix
\section{Filter Function Calculation}\label{app:filterFunction}

The filter transfer function $F(\omega)$ is calculated according to the procedure outlined in~\cite{Green_2013}, briefly summarised here. The transfer function results from the norm of the \textit{control matrix}, generally introduced to describe the effect that laser noise sources, such as phase and intensity noise, have on the fidelity of a quantum operation. The loss of fidelity of a quantum operation is in fact proportional to the integral of the noise spectrum describing the radiation that induces the operation, averaged through the function $F(\omega)/{\omega}^2$ (Eq.~\ref{eq:chi}). In the particular case of \textit{dephasing} noise, the control matrix is reduced to a three-dimensional row vector ($\vec{R}\left(\omega\right)$), whose squared norm gives $F(\omega)$:
\begin{equation}
    F(\omega) = \left|\vec{R}\left(\omega\right)\right|^2.
    \label{eq:App1}
\end{equation}

The \textit{control vector} $\vec{R}\left(\omega\right)$ for a sequence of pulses is calculated according to the formula,
\begin{equation}
    \vec{R} \left(\omega\right)= \sum_{l=1,..N} e^{i \omega t_{l-1}} \vec{R}^{P_{l}}\left(\omega\right) \Lambda^{(l-1)},
    \label{eq:App2}
\end{equation}

\noindent where the row vector $\vec{R}^{P_{l}} \left(\omega\right)$ is introduced and $\Lambda^{\left(l-1\right)}$ indicates a component, which is a matrix, of the $\vec{\Lambda}$ vector.

In this case of dephasing noise, and for a piecewise defined control sequence, $R^{P_{l}}_{i}\left(\omega\right)$ is given by,

\begin{equation} \label{eq:App3a}
\begin{split}
R^{P_{l}}_{i}\left(\omega\right) & = \frac{\omega}{{\omega}^2-{\Omega_{R}}^2} \Big\{ \delta_{z i} \left ( i \Omega_{R}^{l} g_{l}\left(\omega\right) - \omega f_{l}\left(\omega\right)  \right )  + \\
& + \frac{1}{2} \left ( \Omega_{R}^{l} f_{l}\left(\omega\right) -i \omega g_{l}\left(\omega\right) \right ) \mathrm{Tr} \left(\sigma_{\phi_{l}} \sigma_{z} \sigma_{i}\right) \Big\};
\end{split}
\end{equation}

\begin{subequations} \label{eq:App3}

\noindent where,

\begin{equation} \label{eq:App3b}
    f_{l} (\omega) = \cos{\left(\Omega_{R}^{l} \left(t_{l} - t_{l-1}\right) \right)} e^{i \omega \left( t_{l} - t_{l-1} \right)} - 1,
\end{equation}

\begin{equation} \label{eq:App3c}
    g_{l} (\omega) = \sin{\left(\Omega_{R}^{l} \left(t_{l} - t_{l-1}\right) \right)} e^{i \omega \left( t_{l} - t_{l-1} \right)},
\end{equation}

\begin{equation} \label{eq:App3d}
\sigma_{\phi_{l}} = \sigma_{x} \cos{\phi_{l}} + \sigma_{y} \sin{\phi_{l}},
\end{equation}

\end{subequations}

\noindent where $\delta_{ij}$ is the Kronecker delta function and $\sigma_{i}$ denotes the usual Pauli matrices, with $i=x,y,z$. 

The coefficients of the matrix $\Lambda^{l}$ are then calculated according to,
\begin{subequations}
\begin{align}
    \Lambda^{l}_{i j } & = \frac{1}{2}\,\mathrm{Tr}\left( Q_{l-1}^{\dagger} \sigma_{i} Q_{l-1} \sigma_{j} \right) \label{eq:App4a} ,\\
    Q_{l} & = P_{l} P_{l-1} .. P_{1} \label{eq:App4b} ,\\
    P_{l} & = \exp \left ({- \frac{i}{2} \Omega_{R}^{l} \left(t_{l} - t_{l-1}\right) \sigma_{\phi_{l}} } \right ) ,\label{eq:App4c}
    \end{align}
    \label{eq:App4}
\end{subequations}

\noindent where the \textit{cumulative} matrices $Q_{l}$ are introduced and $P_{l}$ is the propagator of the $l$-th operation of a quantum sequence. Observe that for the case $l=0$: $Q_{0}=P_{0}=\mathcal{I}$. 

The calculation of the transfer function therefore requires the definition of a chain of $N$ operations ($l=1,..N$) of \textit{pseudo-rotation}, sequenced through a vector of operation times $\vec{t} = \{t_{l}\,|\, l=1,..N\}$, a vector of Rabi frequencies $\{\Omega_{R}^{l}\}$, describing the rotations' amplitudes, and a vector of azimuthal angles $\{\phi_{l}\}$ describing the rotations' axes, with $\phi=0$ for a rotation about the $x$ axis and $\phi=\pi$ for a counter-propagating beam. For the case of free propagation, $\Omega_{R}^{l}$=0 and $P_{l}=\mathcal{I}$.

The required sum on $l$ described in Eq.~\ref{eq:App2} requires the calculation of N functions, each obtained by performing the calculation described by Eqs.~\ref{eq:App3a}-\ref{eq:App4} relatively to step $l$, which must be iterated for values $l=1,..N$.

% Bibliography
\bibliography{main}

\end{document}